\documentclass[aps,twocolumn,showpacs]{revtex4}

\usepackage{graphicx}

\newcommand{\beq}{\begin{equation}} \newcommand{\eeq}{\end{equation}}
\newcommand{\beqa}{\begin{eqnarray}} \newcommand{\eeqa}{\end{eqnarray}}
 \newcommand{\w}{\omega}

\newcommand{\ket}[1]{\left| #1 \right\rangle}
\newcommand{\bra}[1]{\left\langle #1 \right|}

\begin{document}

\title{Creating excitonic entanglement in quantum dots through the
optical Stark effect}
\author{Ahsan~Nazir}
\email{ahsan.nazir@materials.ox.ac.uk} \affiliation{Department of
Materials, Oxford University, Oxford OX1 3PH, United Kingdom}
\author{Brendon~W.~Lovett} \email{brendon.lovett@materials.ox.ac.uk}
\affiliation{Department of Materials, Oxford
University, Oxford OX1 3PH, United Kingdom}
\author{G.~Andrew~D.~Briggs} \affiliation{Department of Materials,
Oxford University, Oxford OX1 3PH, United Kingdom}
\date{\today}
\begin{abstract}
We show that two initially non-resonant quantum dots may be brought into resonance by the application of a
single detuned laser. This allows for control of the inter-dot interactions and the generation of highly entangled
excitonic states on the picosecond timescale. Along with arbitrary single qubit manipulations, this system would be
sufficient for the demonstration of a prototype excitonic quantum computer.
\end{abstract}

\pacs{03.67.Mn, 78.67.Hc}

\maketitle

Semiconductor quantum dots (QD's) are often described as `artificial
atoms' due to the discrete energy level structure which results from
their three-dimensional confinement~\cite{jacak98}. Consequently, many
of the techniques of quantum optics are now being used in QD
studies and have led to the observation of Rabi oscillations~\cite{stievater01},
photon antibunching~\cite{michler00}, and recently the optical Stark effect~\cite{unold04} in single QD's.
Such experiments have stimulated a great deal of interest in possible
applications of QD's in quantum information processing (QIP) devices~\cite{quiroga99,biolatti02,troiani00}.

In this paper, we shall analyze the behaviour of two adjacent
self-assembled QD's addressed by an external classical laser field, with
the aim of controlling the electronic interactions between them. We shall demonstrate
that it is possible to generate and maintain long-lived entangled
excitonic states in such QD's through the inter-dot resonant
(F\"orster) energy transfer~\cite{lovett03,quiroga99}. This is achieved
with a single laser that dynamically Stark shifts the exciton ground
states in and out of resonance, effectively switching the inter-dot
interaction on and off.

Our model considers only the ground state (no exciton)
and first excited state (single exciton) in each dot, and these two
states define our qubit as $|0\rangle$ and $|1\rangle$ respectively.
Each QD is assumed to be within the strong-confinement regime where
their typical sizes are much smaller than the corresponding bulk
exciton radius, which is determined by the electron-hole Coulomb interaction. As a result, the confinement energy due to QD size dominates and mixing of the single-particle electron and hole states due to their Coulomb interactions may be
neglected~\cite{schmittrink87}. Any associated energy shift can be absorbed into
the exciton creation energy; this shift is important as
it ensures that the resonance condition for single-particle tunneling
is not the same as that for resonant exciton transfer. Additionally, we consider only weak inter-dot
interaction strengths ($\sim 0.1$~meV) which would be expected for two dots with relatively large
spacing ($\sim 10$~nm)~\cite{nazir04}. Therefore, we may neglect inter-dot tunneling of electrons and holes.

The Hamiltonian for two coupled dots in the presence of a single laser
of frequency $\omega_l$ may be written in the computational basis
$\{|00\rangle,|01\rangle,|10\rangle,|11\rangle\}$ as ($\hbar=1$):
\begin{equation}\label{hamiltonian}
H(t)= \left( \begin{array}{cccc}
\omega_0 & \Omega_2\cos{\omega_l t} & \Omega_1\cos{\omega_l t} & 0 \\
\Omega_2\cos{\omega_l t} & \omega_0+\omega_2 & V_{\rm{F}} &
\Omega_1\cos{\omega_l t} \\   \Omega_1\cos{\omega_l t} & V_{\rm{F}} &
\omega_0+\omega_1 & \Omega_2\cos{\omega_l t} \\   0 &
\Omega_1\cos{\omega_l t} & \Omega_2\cos{\omega_l t} & \omega_{\rm T}+V_{\rm
XX} \\  \end{array} \right),
\end{equation}
where $\omega_0$ is the
ground state energy, $\omega_{1(2)}$ the exciton creation energy for dot $1(2)$, and
$\omega_{\rm T}=\omega_0+\omega_1+\omega_2$. The coupling terms $V_{\rm{F}}$
and $V_{\rm XX}$ are the  F\"orster (transition dipole-dipole) and
biexciton~\cite{troiani00,biolatti02} (static dipole-dipole)
interaction strengths respectively.

We have assumed that each dot may couple to the laser with a different
strength, governed by the respective Rabi frequency $\Omega_1$ or
$\Omega_2$, with
$\Omega_i(t)\equiv-2{\bf d}_i\cdot{\bf E}({\bf r},t)$,
for $i=1,2$. Here, ${\bf d}_i$ is the inter-band ground state transition dipole moment for dot
$i$, and ${\bf E}({\bf r},t)$ is the laser amplitude at time $t$ and
position ${\bf r}$. Natural size and composition fluctuations in
self-assembled dot samples (for example in InGaAs QD's~\cite{eliseev00}) lead to
a large range of possible transition dipole moments for each dot.
The size of the ground-state dipole mismatch (and related exciton energy difference) between
two dots is an important factor in determining our ability to
control their interactions.

We shall first analyze the Hamiltonian of Eq.~\ref{hamiltonian} within
the rotating wave approximation (RWA). This will allow us to derive
approximate conditions governing the behaviour of the system, and
elucidate the mechanism for controlling excitonic entanglement.
Subsequently, we shall characterize the small corrections to these RWA
solutions and perform a full numerical solution of $H(t)$.

Transforming Eq.~\ref{hamiltonian} into a frame rotating with the laser
frequency $\omega_l$ with respect to both qubits, we obtain (within the
RWA): \begin{equation}\label{Hrotating}
H'=\left( \begin{array}{cccc}
 0 & \Omega_2/2 & \Omega_1/2 & 0 \\   \Omega_2/2 & \delta_2 & V_{\rm F}
& \Omega_1/2 \\   \Omega_1/2 & V_{\rm F} & \delta_1 & \Omega_2/2 \\   0
& \Omega_1/2 & \Omega_2/2 & \delta_1+\delta_2+V_{\rm XX} \\
\end{array} \right),
\end{equation}
where $\delta_i=\omega_i-\omega_l$
is the detuning of the laser from dot $i$, and an irrelevant constant
has been subtracted from each term on the diagonal. In order to
demonstrate control over the interaction $V_{\rm F}$ we would like to
isolate the behaviour of the $\{|01\rangle,|10\rangle\}$ subspace in
which it acts. We may proceed, utilizing degenerate perturbation theory,
provided that the following conditions are satisfied:
\begin{equation}\label{conditions}
|\delta_1-\delta_2|,~|V_{\rm F}|,~|\Omega_i/2|\ll min(|\delta_i|,~|\delta_i+V_{\rm
XX}|).
\end{equation}
In this case the three subspaces \{$\ket{01}$,$\ket{10}$\}, \{$\ket{00}$\}, and \{$\ket{11}$\}
are decoupled and we can write an effective Hamiltonian for the
$\{|01\rangle,|10\rangle\}$ subspace:
\begin{equation}\label{0110}
\left( \begin{array}{cc}   \delta_2+\alpha\Omega_2'^2-\beta\Omega_1'^2
& V_{\rm F}   +\Omega_1'\Omega_2'(\alpha-\beta) \\   V_{\rm
F}+\Omega_1'\Omega_2'(\alpha-\beta) &
\delta_1+\alpha\Omega_1'^2-\beta\Omega_2'^2 \\  \end{array} \right),
\end{equation} where $\alpha=1/\delta_1$, $\beta=1/(\delta_2+V_{\rm
XX})$, and $\Omega_i'=\Omega_i/2$.

\begin{figure}[t] \centering
\includegraphics[width=2.8in,height=2.8in]{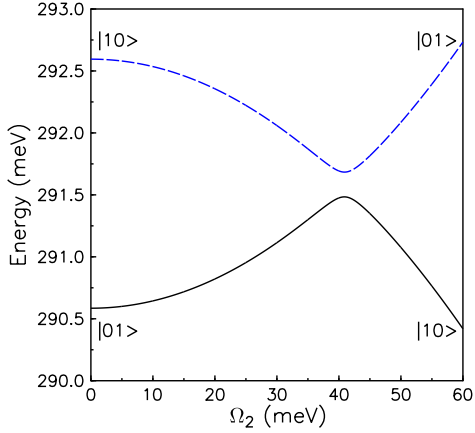}
\caption{Laser induced anticrossing in the $\{|01\rangle,|10\rangle\}$ subspace, for fixed ratio $\Omega_1/\Omega_2=0.55$. Eigenvalues are calculated from Eq.~{\ref{0110}} with $\delta_1=292.59$~meV, $\delta_2=290.59$~meV, and $V_{\rm F}=0.1$~meV.}
\label{laseracross} \end{figure}

As we may control both the detunings $\delta_i$ of the laser from the
QD's, and the Rabi frequencies $\Omega_i$, it
is possible to controllably modify the dynamics within this subspace.
As is shown in Fig.~\ref{laseracross}, two regimes are of particular
interest. When the laser is off, the difference in diagonal elements can be much larger than the effective interaction strength if
\begin{equation}\label{nonresonant}
\w_1-\w_2\gg V_{\rm F} .
\end{equation}
In this case, the eigenstates of Eq.~\ref{0110} approach the computational basis states $|01\rangle$ and
$|10\rangle$, shown away from the anticrossing in Fig.~{\ref{laseracross}}, that would be expected for
non-interacting dots.

In contrast, under the condition:
\begin{equation}\label{resonance}
\delta_1-\delta_2=\omega_1-\omega_2=\left[\Omega_2'^2-\Omega_1'^2\right](\alpha+\beta),
\end{equation}
the diagonal terms of Eq.~\ref{0110} are
equal and the two dots Stark shift into resonance under the action of
the laser. The corresponding eigenstates lie at the anticrossing and are maximally entangled due to
the modified off-diagonal interaction
\begin{equation}
V_{\rm eff}\equiv V_{\rm F}+\Omega_1'\Omega_2'(\alpha-\beta).
\end{equation}
They are given by $|\psi_+\rangle=2^{-1/2}(|01\rangle+|10\rangle)$ and
$|\psi_-\rangle=2^{-1/2}(|10\rangle-|01\rangle)$. Hence, if the system is
initialized in the state $|01\rangle$, it will coherently evolve to
$i|10\rangle$ during the laser pulse, passing through the maximally
entangled state $2^{-1/2}(|01\rangle+i|10\rangle)$. This happens
with a coherent exciton transfer time of $t=\pi/(2V_{\rm eff})$.

Therefore, we may selectively couple the two initially non-resonant
QD's (satisfying Eq.~\ref{nonresonant} before the laser is switched on)
by the application of a single detuned laser satisfying Eq.~\ref{resonance}
which non-adiabatically shifts the eigenstates to the anticrossing point in Fig.~{\ref{laseracross}}.
As soon as we wish to decouple the dots again, we simply turn the laser off.

\begin{figure}[t] \centering
\includegraphics[width=3.4in,height=3.5in]{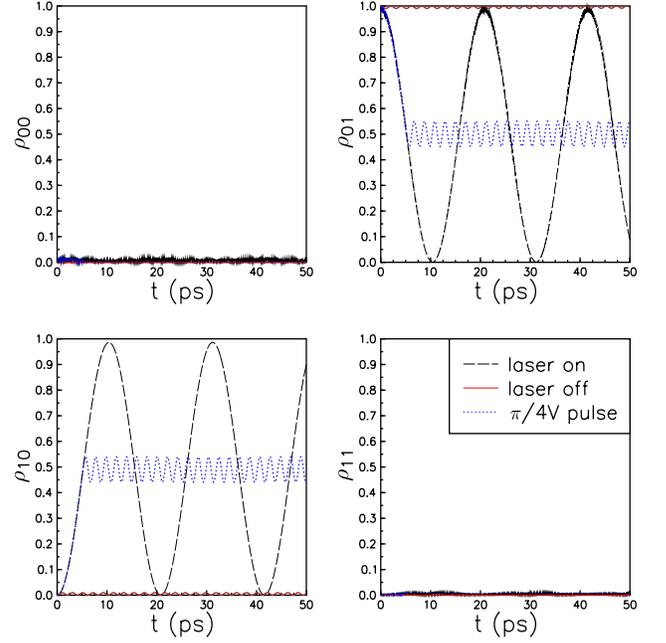}
\caption{Populations of the four states $\ket{00}, \ket{01}, \ket{11}$ and $\ket{10}$
calculated from Eq.~\ref{Hrotating} with input $|01\rangle$,
$\rho_{nm}=\langle nm|\rho|nm\rangle$, $V_{\rm F}=0.1$~meV and $V_{\rm XX}=0$.
Dashed line: state evolution when the laser is always on. The
Rabi frequencies and detunings of the laser are given by $\Omega_2~=~40.96$~meV,
$\Omega_1=0.55\Omega_2$, $\Omega_2/2\delta_1=0.07$, and
$\delta_1-\delta_2=2$~meV. Dotted line:
state evolution when the same laser is on for a time of $\pi/(4V_{\rm eff})$ and then turned off.
The small oscillations in population in the $\pi/(4V_{\rm eff})$ pulse case after the laser is
switched off are due to some residual coupling between
the dots which can be suppressed by increasing the energy selectivity
$\delta_1-\delta_2$.
Solid line: state evolution when the laser is always off. In this case the state is almost purely $\ket{01}$
throughout.} \label{rwa} \end{figure}

This effect is demonstrated in Fig.~\ref{rwa} where a numerical
simulation of the evolution of an input state $|01\rangle$ under the
RWA Hamiltonian of Eq.~\ref{Hrotating} is shown. Without laser
coupling, the system remains in its initial state $|01\rangle$ with a
fidelity, ${\cal F}>1-4\left(V_{\rm F}/|\delta_1-\delta_2|\right)^2=0.99$.
However, under the application
of a laser satisfying Eq.~\ref{resonance}, coherent oscillations
are observed between the states $|01\rangle$ and $|10\rangle$ with an
exciton transfer time of $10.9$~ps, for the parameters
chosen here. This transfer is very close to being $100\%$ complete with
little population leaking from the $\{|01\rangle,|10\rangle\}$
subspace, justifying our perturbative treatment.

This behaviour could be observed
in an experiment by applying a laser pulse for a series of different pulse lengths, and afterwards
observing the emitted photons. As can be seen in Fig.~\ref{laseracross}, the two eigenstates
when the laser is off (which are approximately $\ket{10}$ and $\ket{01}$) are separated by some 2~meV. This is readily
resolved in a modern spectrometer -- and so
a measurement of the wavelength of the emitted photons can be use to determine which of the two dots each one came from.
Plotting the number of each wavelength of detected photons as a function of the pulse length
would allow a determination of the coherent transfer oscillations.
Further, a $t=\pi/(4V_{\rm eff})$ pulse will create and maintain an
(approximately maximally) entangled state from an initially separable
one, and this is also shown in Fig~\ref{rwa}. Even for the small coupling
strength ($V_{\rm F}=0.1$~meV) considered here this
operation is on the picosecond time scale. Therefore, we would expect
such an entangled state to be long-lived in a pair of QD's
relative to the timescale of its generation. Single and coupled dot
exciton lifetimes have been measured to be as long as nanoseconds at
low temperatures~\cite{bayer02,birkedal01,borri03}. Additionally, pure phonon dephasing
effects are suppressed as the temperature is lowered below 10~K~\cite{bayer02,borri01}.

\begin{figure}[t] \centering
\includegraphics[width=3.0in,height=5.0in]{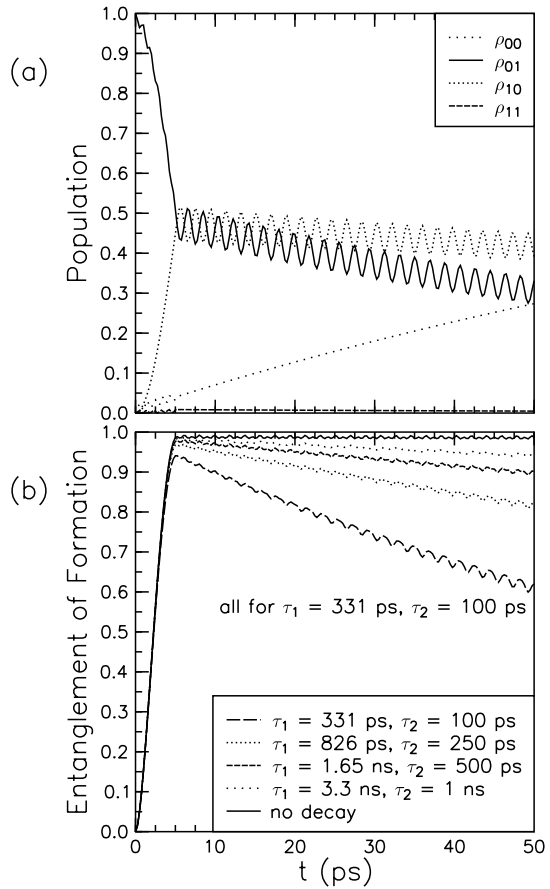}
\caption{(a) Numerical simulation of the four populations
for an input state $\ket{01}$ and a $\pi/(4V_{\rm eff})$ pulse. The simulation uses
Eq.~\ref{mastereq}, which does not rely on the RWA and includes exciton decay,
but not pure dephasing.  The decay rates for dots $1$ and $2$ are given by $\Gamma_1 =\tau_1^{-1} = (331~{\rm ps})^{-1}$ and
$\Gamma_2=\tau_2^{-1} = (100~{\rm ps})^{-1}$ respectively.
All other parameters are the same as for
Fig.~\ref{rwa} except $\delta_1-\delta_2=2.18$~meV to account for the extra shifts, and $\omega_l=1500$~meV.
(b) Entanglement of formation of the input state $|01\rangle$ as a function of time
for a $\pi/(4V_{\rm eff})$ laser pulse, and for a series of different decay rates.
We keep a constant ratio of $\Gamma_1/\Gamma_2=|{\bf d}_1/{\bf d_2}|^2=(\Omega_1/\Omega_2)^2$.
The calculations are made by using Eqs.~\ref{mastereq}-\ref{tau}.
}
\label{eof} \end{figure}

We therefore now account for the finite exciton lifetimes by including only spontaneous
emission terms in the density matrix master equation~\cite{agarwal74}, and neglecting pure dephasing processes:
\begin{equation}\label{mastereq}
\dot{\rho}=-i[H,\rho]+\frac{1}{2}\sum_{i}\Gamma_{i}
\left(2\sigma_{i}^-\rho\sigma_{i}^+-\sigma_{i}^+\sigma_{i}^-\rho-\rho\sigma_{i}^
+\sigma_{i}^-\right).
\end{equation}
Here, the $i$ label the dipole allowed transitions in the coupled system,
$\sigma_{i}^+$ and $\sigma_{i}^-$ are their raising and lowering operators, and the $\Gamma_i$ are
their transition rates.
These terms can lead to a significant reduction of the degree of entanglement over time.
We can see this by referring to Fig.~\ref{eof}, where we show the result of numerical calculations which use the full
Hamiltonian (Eq.~\ref{hamiltonian}) in the master equation (Eq.~\ref{mastereq}).
In Fig.~\ref{eof} (a) the populations of the four states are shown
as a function of time for the input state
$|01\rangle$, subject to a square pulse of $5.45$~ps duration (this satisfies $t=\pi/(4V_{\rm eff})$
for our chosen parameters), and subject to significant decay. In Fig.~\ref{eof} (b),
we plot the entanglement of formation (EOF)~\cite{bennett96} of the system
as a function of time for the input state
$|01\rangle$, for a variety of decay rates.
The EOF measures the number of Bell states required to create
the state of interest; for a maximally entangled state it is equal to unity
while for a separable state it is zero. For a general two qubit state it is given by the equation
\begin{equation}
E_F({\rho})=h\left(\frac{1+\sqrt{1-\tau}}{2}\right),
\end{equation}
where $h(x) = -x\log_2(x) - (1 - x)\log_2(1 - x)$ is the Shannon
entropy function. $\tau$ is the ``tangle'' or ``concurrence'' squared, which can be computed by
using the equation:
\begin{equation}
\tau={\cal
C}^{2}=\left[\max\{\lambda_1-\lambda_2-\lambda_3-\lambda_4,0\}\right]^{2}.
\label{tau}
\end{equation}
Here the $\lambda$'s are the square roots of the eigenvalues, in
decreasing order, of the matrix
${\rho} \tilde{{\rho}} = {\rho}\;\sigma_{y}^{A}
\otimes \sigma_{y}^{B} {\rho}^{*} \sigma_y^A \otimes
\sigma_{y}^{B}$, where ${\rho}^{*}$ denotes the complex conjugation
of ${\rho}$
in the computational basis $\ket{00}, \ket{01}, \ket{10}, \ket{11}$~\cite{munro01}.
In Fig.~\ref{eof} we see that at the end of the laser pulse the state has become almost
maximally entangled and, in the absence of decay, stays so once the laser is switched off (the small deviation from unity
is due to the residual effect of the F\"orster coupling when the laser is off).
When decay is included the EOF decreases over time,
but for typical experimental lifetimes of 1 ns~\cite{bayer02,birkedal01,borri03},
it retains a value which is higher than 0.94 for times up to 50~ps~\cite{others}.

The numerical solution of Eq.~\ref{mastereq} required for Fig.~\ref{eof} has been computed without making use of
the RWA. The behaviour of the system is exactly as expected from the RWA case, except for an extra small shift in the dot
energies when the laser is on. This shift is the largest correction to the Stark shift in a perturbative expansion and
arises from the counter-propagating terms in the
state evolution which are discarded when the RWA is made~\cite{barnett97}.
The corrections to the RWA can all be derived by following the method of Shirley~\cite{shirley65}, who considers the case
of a two-level system (which we label 0 and 1 and consider to have an energy separation of $\w_1$)
coupled by an oscillating field (of frequency $\w_l$). He shows that this kind of problem can be mapped on to a
time independent one by using Floquet's theory to construct a Hamiltonian in an infinite dimensional Hilbert space
given by
\beq
\bra{\alpha n} {\cal H}_F \ket{\beta m} = {\cal H}_{\alpha\beta}^{n-m} + n\w \delta_{\alpha\beta}\delta_{nm}.
\eeq
Here $\alpha, \beta \in \{0, 1\}$ and the $n, m$ represent Fourier components of the state evolution.
${\cal H}^n_{\alpha\beta}$ is the Fourier
component with frequency $n\w$ of the oscillating Hamiltonian (which is only non-zero for $\alpha \neq \beta$).
Shirley showed that the time evolution operator
of the two-level system can be written as
\beq
U_{\beta\alpha}(t;t_0) = \sum_n \bra{\beta n}\exp[-i{\cal H}_F(t-t_0)]\ket{\alpha 0} e^{in\w t}.
\label{tevol}
\eeq
He goes on to derive the Bloch-Siegert shift, which is important for resonant interactions between the
two-level system and the oscillating field. However, we are interested in the off-resonant behaviour. In the absence of any
interactions, our system will initially be in a state of the form
$a\ket{0 0}+b\ket{1 0}$ (in the notation $\ket{\alpha n}$), and
we can use the time evolution operator (Eq.~\ref{tevol}) to determine the energy separation of the two
(eigen)states, $\ket{0 0}$ and $\ket{1 0}$. We follow a similar procedure when the laser is switched on.
In this case, if the laser is sufficiently detuned from resonance (i.e. if
$\delta = (\w_1  - \w_l) \gg \Omega/2$, with $\Omega$ the laser-system coupling),
the $\ket{\alpha n}$ are still approximate
eigenstates and we can employ second order perturbation theory to obtain the energy
shift in the separation of the two levels.
We obtain:
\beq
\Delta = \frac{2\Omega^{\prime 2}}{\delta} + \frac{2\Omega^{\prime 2}}{(\delta + 2\w_l)}
\eeq
where, as before, $\Omega^\prime = \Omega/2$.
The extra term represents a detuning
from resonance by an amount $2\omega_l+\delta$, compared to the
term with detuning $\delta$ which is usually the only one kept. For the
relatively large detunings considered in our two dot model, they become non-negligible
contributions to the Stark shifts for each dot, $i$, with a magnitude of
$2\Omega_i'^2/(2\omega_l+\delta_i)$. Once we account for this extra shift, for example in this calculation by redefining
the parameters $\delta_1$ and $\delta_2$ to include it, the system behaves exactly as expected from our earlier
analysis of Eq.~\ref{0110}.

We now assess the feasibility of our method by examining the state-of-the-art in real systems.
In our previous work~\cite{lovett03}, we predicted that the F\"orster transfer
energy can be as large as about 1~meV in QD's (corresponding to energy transfer times on the
sub-picosecond timescale); we also showed how a static electric field can be applied to the dots to suppress
the interaction and so lengthen the transfer time as required.
In addition to this, experimental work has suggested that F\"orster transfer times can
approach picoseconds in structurally optimized pairs of CdSe QD's~\cite{crooker02}, and that
these transfer times can
be on the subpicosecond timescale in photosynthetic biomolecular systems~\cite{herek02}. Molecular systems could
therefore provide an alternative route to experimental realization of the effects we have predicted here.

QD exciton - laser interaction strengths (which correspond to the Rabi frequency when the laser is resonant
with the exciton) of a few meV have been attained in an experiment which measured the optical Stark shift in AlAs/GaAs
heterostructures~\cite{unold04}. Further, an experiment which directly observed Rabi oscillations in a InGaAs/GaAs
QD photodiode also measured a Rabi frequency of a similar magnitude~\cite{zrenner02}.
However, these experiments were not designed to maximize
the laser - exciton coupling strength, and so the value of 40 meV, which we used in our simulations,
is not unrealistic.

The coherent exciton transfer process detailed in Fig.~\ref{rwa} is
equivalent to the realization of iSWAP logic operations between the pair of excitonic qubits~\cite{schuch03}.
Along with arbitrary single qubit rotations, this gate would be sufficient
for a demonstration of a prototype excitonic quantum computer.
Single qubit operations in our two qubit system may be achieved by using their frequency addressability. A laser resonantly
tuned to either of the two dots will induce a Rabi oscillation in the dot to which it is tuned; in the language of
Pauli spin matices this is a rotation around the $X$ axis of the Bloch sphere, and represents one of the two
rotations which are required for arbitrary single qubit operations. The other could be obtained in a number of ways.
For example
a slightly detuned laser will cause the qubit state to move around another trajectory on the Bloch sphere.
Alternatively,
a higher level $\ket{T}$ (of energy $\omega_T$ relative to the $\ket{0}$) could be used.
If this higher level has a different energy for each qubit, it could be resonantly
excited from, say, the $\ket{0}$ state in only one of the two qubits by applying an appropriately tuned $\pi$ laser pulse.
Leaving the system to evolve then for a time $\tau$ before using another $\pi$ pulse to deexcite it causes the $\ket{0}$
to pick up a phase of $\w_T\tau$ relative to the $\ket{1}$. This $Z$ rotation is
sufficient to complete the set required for universal quantum computing.

If the ratio of decoherence time to gate operation time
in our system were large enough, full scale fault tolerant quantum computing (FTQC)~\cite{steane03} would be possible.
It is well established that a ratio of around 1000 is
good enough for FTQC, and recent estimates of this have been a low as 100~\cite{reichardt04}.
We have demonstrated that an entangling gate
can be performed in real systems in around 5~ps, which is around 200 times shorter than the longest measurements of
decoherence
times~\cite{bayer02}. Hence, our proposed scheme could be carried out with a precision which
is already on the limit required for FTQC; incremental improvements in either FTQC protocols or in experimental
systems should enable the implementation of full quantum algorithms.
The absolute speed at which gates can be carried out in our scheme also makes them ideal for smaller scale
applications
such as quantum repeaters~\cite{dur99}, which require much less stringent gate fidelities than full scale FTQC.

To summarize, we have shown that two non-resonant QD's may be
brought into resonance by the application of a single detuned laser
which induces Stark shifts within each dot without significant population excitation.
This in turn allows for control over the inter-dot interactions, and
hence the generation of highly entangled states on the picosecond
timescale. The conditions of Eqs.~\ref{conditions},~\ref{nonresonant}, and~\ref{resonance} set the upper
limits on the energy selectivity and, neglecting incoherent processes, the fidelity possible with a particular dot sample.
The F\"{o}rster strength $V_{\rm F}$ sets the interaction timescale. In general, as the magnitude and difference of the
Rabi frequencies increases, and as $V_{\rm F}$ increases, so does the feasibility of the proposed idea.
We believe that the means to demonstrate the effects outlined above already exist.
Moreover, these types of experiments may prove invaluable in assessing the potential applicability of
semiconductor QD's for future QIP technologies.

AN and BWL are supported by EPSRC (BWL as part of the Foresight LINK
Award ``Nanoelectronics at the Quantum Edge," GR/R66029, and the QIPIRC, GR/S82176).
We thank T. P. Spiller, S. D. Barrett, and W. J. Munro for useful and stimulating
discussions, and R. G. Beausoleil for providing numerical simulation code.


\begin{thebibliography}{23}
\bibitem[{\citenamefont{Reichardt}()\citenamefont{others}}]{others}
\bibinfo
  \eprint{If $\ket{00}$ or $\ket{11}$ are used as the input states, $E_{F} = 0$ throughout the subsequent evolution.}.
\end{thebibliography}
\end{document}